\documentclass[12pt]{article}

\usepackage{amsmath}
\usepackage{amssymb}
\usepackage{amsfonts}
\usepackage{latexsym}
\usepackage{bm}
\usepackage{color}

\catcode `\@=11 \@addtoreset{equation}{section}
\def\theequation{\arabic{section}.\arabic{equation}}
\catcode `\@=12

\newcommand{\be}{\begin{equation}}
\newcommand{\en}{\end{equation}}
\newcommand{\bea}{\begin{eqnarray}}
\newcommand{\ena}{\end{eqnarray}}
\newcommand{\beano}{\begin{eqnarray*}}
\newcommand{\enano}{\end{eqnarray*}}
\newcommand{\bee}{\begin{enumerate}}
\newcommand{\ene}{\end{enumerate}}

\newcommand{\mc}{\mathcal}

\newcommand{\Sc}{{\cal S}}
\newcommand{\E}{{\cal E}}
\newcommand{\F}{{\cal F}}

\newcommand{\Lc}{{\cal L}}

\newcommand{\1}{1 \!\! 1}

\newcommand{\Hil}{\mc H}

\catcode `\@=11 \@addtoreset{equation}{section}
\catcode `\@=12

\begin{document}

\thispagestyle{empty}

\vspace*{2cm}

\begin{center}
{{\Large \bf Non self-adjoint Hamiltonians with complex eigenvalues}}\\[10mm]


{\large F. Bagarello} \footnote[1]{ Dipartimento di Energia, Ingegneria dell'Informazione e Modelli Matematici,
Facolt\`a di Ingegneria, Universit\`a di Palermo, I-90128  Palermo, and INFN, Universit\`a di di Torino, ITALY\\
e-mail: fabio.bagarello@unipa.it\,\,\,\, Home page: www.unipa.it/fabio.bagarello}

\end{center}

\vspace*{2cm}

\begin{abstract}
\noindent  Motivated by what one observes dealing with PT-symmetric quantum mechanics, we discuss what happens if a physical system is driven by a diagonalizable Hamiltonian with not all real eigenvalues. In particular, we consider the functional structure related to  systems living in finite-dimensional Hilbert spaces, and we show that certain intertwining relations can be deduced also in this case if we introduce suitable antilinear operators. We also analyze a simple model, computing the transition probabilities in the broken and in the unbroken regime.

\end{abstract}


\vfill


\newpage

\section{Introduction}

In ordinary quantum mechanics one of the fundamental axiom of the whole theory is that the Hamiltonian $H$ of the physical system must be self-adjoint: $H=H^\dagger$. This condition, shared also by all the {\em observables} of the system, is important since it ensures that the eigenvalues of these observables, and of the hamiltonian in particular, are real quantities. However, this is not a necessary condition, and in fact several physically motivated examples can be found in the literature considering non self-adjoint operators whose spectra consist of only real eigenvalues, see \cite{baginbagbook} and references therein, for instance.  This situation is particularly interesting for people in the PT-community, who quite often work with Hamiltonian operators which are not self-adjoint, but simply pseudo-symmetric or PT-symmetric, \cite{ben,ali}. In two recent papers we have investigated Hamiltonians which are not self-adjoint, but still possess real eigenvalues, mainly with the aim of understanding how the dynamics and the transition probability should be defined, or at least postulated, in these cases. This has been done first in finite-dimensional and later in infinite dimensional Hilbert spaces, \cite{bagaop1} and \cite{bagaop2}. Already in \cite{bagaop1} the importance of having only real eigenvalues was pointed out: going from real to complex eigenvalues of $H$ can, in fact be very unpleasant from a purely abstract point of view.  However, in PT quantum mechanics, this is exactly the kind of transition which is of some interest in concrete situations, and for this reason it is definitely interesting, in our opinion, to carry on a general analysis of this case. This is what this paper is about: Hamiltonians with (some) complex eigenvalues, and some possible applications of these operators.

All throughout this paper we will work with finite-dimensional Hilbert spaces. Hence, all the operators involved in our analysis are bounded (hence, everywhere defined) and their inverse, when they exist, are bounded as well. Of course, going from a finite to an infinite dimensional Hilbert space is an absolutely non trivial task. Therefore, most of our claims, though giving indications also in this latter case, are rigorously true only in the present, finite-dimensional, settings. We will comment more on this aspect all along the paper.

This article is organized as follows:

In the next section, after a brief review of what has been done in \cite{bagaop1}, we consider which kind of new features arise  when the Hamiltonian possesses non real eigenvalues. In particular, we discuss the role of antilinear operators in this context and we show how this type of operators are needed for deducing useful intertwining relations. In this way we will be able to find isospectral operators which are constructed out of $H$ and $H^\dagger$. Notice that, of course, $H$ and $H^\dagger$ cannot be isospectral by themselves, due to the presence of complex eigenvalues. Section 3 is devoted to an example taken from the literature, while our conclusions are contained in Section 4. To keep the paper self-consistent,  in the Appendix we give some results on antilinear operators which are used all throughout the paper and cannot be easily found in the literature.

\section{A general settings for $H\neq H^\dagger$}

As we have already said, in this paper we will focus on the easiest situation, i.e. on finite dimensional Hilbert spaces. In this way our linear operators are finite matrices. The main ingredient is a linear operator (i.e. a matrix) $H$, acting on the vector space ${\Bbb C}^{N+1}$, with $H\neq H^\dagger$ and with exactly $N+1$ distinct eigenvalues $E_n$, $n=0,1,2,\ldots,N$. Here, the adjoint $H^\dagger$ of $H$ is the usual one, i.e. the complex conjugate of the transpose of the matrix $H$. Because of what follows, and in order to fix the ideas, it is useful to remind here that the adjoint of an operator $X$, $X^\dagger$, is defined in terms of the {\em natural} scalar product $\left<.,.\right>$ of the Hilbert space $\Hil=\left({\Bbb C}^{N+1},\left<.,.\right>\right)$:
$\left<Xf,g\right>=\left<f,X^\dagger g\right>$, for all $f,g\in {\Bbb C}^{N+1}$, where $\left<f,g\right>=\sum_{k=0}^N\overline{f_k}\,g_k$, with obvious notation. As stressed in the Introduction, our main interest here is to the case when some of the eigenvalues $E_n$ of $H$ are not real. However, to simplify the treatment, we will assume that each eigenvalue $E_n$ has multiplicity one. The extension to higher finite multiplicities is easy, but it will not be considered here to avoid making the notation unnecessarily heavier.

Before starting, it is necessary to clarify some notation adopted in this paper: we will use ${\Bbb C}^{N+1}$ any time we only want to stress the nature of vector space of the set of our vectors. When it is important to stress the topological (i.e. the scalar products and the norms) aspects of this set, we will use $\Hil$ instead of ${\Bbb C}^{N+1}$ (and, later, $\Hil_\varphi$ or $\Hil_\Psi$). Before starting with our analysis, we recall few of our results in \cite{bagaop1}, in order to see later the differences with the situation which is more relevant for us in this paper, i.e. the case in which the imaginary part of some eigenvalue of $H$ is different from zero.

\subsection{All the eigenvalues are real: a review}\label{sectRE}

We assume here that $H$ has $N+1$ distinct real eigenvalues, corresponding to $N+1$ eigenvectors $\varphi_k$, $k=0,1,2,\ldots,N$:
\be
H\varphi_k=E_k\varphi_k.
\label{21}\en
The set $\F_\varphi=\{\varphi_k,\,k=0,1,2,\ldots,N\}$ is a basis for ${\Bbb C}^{N+1}$, since the eigenvalues are all different. Then an unique biorthogonal basis of $\Hil$, $\F_\Psi=\{\Psi_k,\,k=0,1,2,\ldots,N\}$, surely exists, \cite{you,chri}: $\left<\varphi_k,\Psi_l\right>=\delta_{k,l}$, for all $k, l$, and we have $H^\dagger\Psi_k=E_k\Psi_k$ for all $k$.  Moreover, for all $f\in\Hil$, we can write
$f=\sum_{k=0}^N\left<\varphi_k,f\right>\Psi_k = \sum_{k=0}^N\left<\Psi_k,f\right>\varphi_k$. Hence, both $\F_\varphi$ and $\F_\Psi$ are complete (or total): if $f\in\Hil$ is such that $\left<\varphi_k,f\right>=0$, or $\left<\Psi_k,f\right>=0$, for all $k$, then $f=0$.

Using the bra-ket notation we can write $\sum_{k=0}^N|\varphi_k\left>\right<\Psi_k|=\sum_{k=0}^N|\Psi_k\left>\right<\varphi_k|=\1$, where, for all $f,g,h\in\Hil$, we define $(|f\left>\right<g|)h:=\left<g,h\right>f$. Introducing, as usual, the operators $S_\varphi=\sum_{k=0}^N|\varphi_k\left>\right<\varphi_k|$ and $S_\Psi=\sum_{k=0}^N|\Psi_k\left>\right<\Psi_k|$, we know that these are bounded positive, self-adjoint, invertible operators, one the inverse of the other: $S_\Psi=S_\varphi^{-1}$. We want to stress  that, in our present settings, there is absolutely no problem with the domains of these (and other) operators, while in \cite{baginbagbook} and in \cite{bagaop2} we have discussed what happens for infinite-dimensional Hilbert spaces. This generalization is absolutely non trivial, but will not be considered here. Using standard techniques in functional analysis, or direct matrix computations, we can introduce the positive square roots of $S_\Psi$ and $S_\varphi$, and again we have $S_\Psi^{1/2}=S_\varphi^{-1/2}$. Useful (and well known) properties are the following:
\be
S_\varphi\Psi_n=\varphi_n,\quad S_\Psi\varphi_n=\Psi_n,\quad \mbox{ as well as }\quad S_\Psi H=H^\dagger S_\Psi,\quad S_\varphi H^\dagger= HS_\varphi.
\label{23}\en
 If we now define $H_0:=S_\Psi^{1/2}HS_\varphi^{1/2}$ and $e_k=S_\Psi^{1/2}\varphi_k= S_\varphi^{1/2}\Psi_k$, $k=0,1,2,\ldots,N$, we see that $H_0=H_0^\dagger=S_\varphi^{1/2}H^\dagger S_\Psi^{1/2}$,  and that $\E=\{e_k,\,k=0,1,2,\ldots,N\}$ is an orthonormal (o.n.) basis of $\Hil$ of eigenstates of the self-adjoint operator $H_0$: $H_0e_k=E_ke_k$.

Similarly to what is done in many places in the literature, $S_\Psi$ and $S_\varphi$ can be used now to define new scalar products in ${\Bbb C}^{N+1}$:
\be
\left<f ,g\right>_\varphi:=\left<S_\varphi^{1/2}f ,S_\varphi^{1/2}g\right>=\left<S_\varphi f ,g\right>, \qquad
\left<f ,g\right>_\Psi:=\left<S_\Psi^{1/2}f ,S_\Psi^{1/2}g\right>=\left<S_\Psi f ,g\right>,
\label{24}\en
for all $f,g\in{\Bbb C}^{N+1}$. Due to the properties of $S_\Psi$ and $S_\varphi$, these are really scalar products, everywhere defined on ${\Bbb C}^{N+1}$. Of course, the related norms $\|.\|$, $\|.\|_\varphi$ and $\|.\|_\Psi$ are all equivalent\footnote{This means that, if a sequence of vectors $f_n\in {\Bbb C}^{N+1}$ converges in $\|.\|$, it also converges in $\|.\|_\varphi$ and in $\|.\|_\Psi$.}, since, for instance
$$
\frac{1}{\|S_\varphi^{-1/2}\|}\,\|f\|\leq\|f\|_\varphi\leq \|S_\varphi^{1/2}\|\,\|f\|,
$$
for all $f\in{\Bbb C}^{N+1}$. A similar double inequality could be deduced also for $\|f\|_\Psi$. Then, from a topological point of view, $\Hil$, $\Hil_\varphi:=\left({\Bbb C}^{N+1},\left<.,.\right>_\varphi\right)$ and $\Hil_\Psi:=\left({\Bbb C}^{N+1},\left<.,.\right>_\Psi\right)$ are all equivalent. However, they are different under other aspects: to begin with, they differ in the definition of the adjoint of the operators, which is $\dagger$ in $\Hil$, but which becomes $\flat$ in $\Hil_\varphi$ and $\sharp$ in $\Hil_\Psi$: $\left<Xf,g\right>_\varphi=\left<f,X^\flat g\right>_\varphi$ and $\left<Xf,g\right>_\Psi=\left<f,X^\sharp g\right>_\Psi$, for all $f,g\in {\Bbb C}^{N+1}$. It is easy to see that $\sharp$ and $\flat$ are really adjoints\footnote{For instance, $(X^\sharp)^\sharp=X$ and $(XY)^\sharp=Y^\sharp X^\sharp$, for all operators $X$ and $Y$ on ${\Bbb C}^{N+1}$.}, and to deduce the following relations:
\be
X^\flat=S_\Psi X^\dagger S_\varphi,\qquad X^\sharp=S_\varphi X^\dagger S_\Psi\qquad \mbox{and}\quad X^\flat=S_\Psi^2 X^\sharp S_\varphi^2,
\label{2add1}\en
for each operator $X$ on ${\Bbb C}^{N+1}$. It is now an easy computation to check that $H=H^\sharp$, and that $H^\dagger=(H^\dagger)^\flat$. Indeed we have $\left<Hf,g\right>_\Psi=\left<f,Hg\right>_\Psi$ and $\left<H^\dagger f,g\right>_\varphi=\left<f,H^\dagger g\right>_\varphi$, for all $f,g\in {\Bbb C}^{N+1}$. Hence, even if $H$ is not self-adjoint in $\Hil$, $H\neq H^\dagger$, still it turns out to be self-adjoint in a different space, $\Hil_\varphi$. Similarly, $H^\dagger$ is not self-adjoint in $\Hil$, but it is self-adjoint in  $\Hil_\Psi$.

\vspace{2mm}

{\bf Remarks:--} (1) the equalities in (\ref{2add1}) cannot be extended easily if $\dim(\Hil)=\infty$. The reason is the following: if, for instance, $X^\dagger$ and $S_\varphi$ are unbounded, taken $f\in D(S_\varphi)$, the domain of $S_\varphi$, there is no reason a priori for $S_\varphi f$ to belong to $D(X^\dagger)$, so that $X^\dagger S_\varphi f$ needs not to be defined. Moreover, even when this is true, still $X^\dagger S_\varphi f$ is not necessarily a vector in the domain of $S_\Psi$, so that the existence of $S_\Psi X^\dagger S_\varphi f$ is not granted, at least without further assumptions.
 Therefore, in general, when $\dim(\Hil)=\infty$, the three Hilbert spaces are different not only topologically, but also as sets. Of course, this cannot happen if $\dim(\Hil)<\infty$, since all the operators can be defined in all of $\Hil$.

\vspace{1mm}

(2) In \cite{bagaop1}, and later in \cite{bagaop2}, we have used these results in the analysis of the dynamics of a physical system $\Sc$ described by some Hamiltonian with all real and discrete eigenvalues. This analysis is motivated because,  if $H\neq H^\dagger$, some freedom does exist in the definition of the time evolution of $\Sc$, and we have discussed a possible way to remove this freedom, using comparison with experiments and some remarks on the functional structure of the system. We refer to those paper for more results on these aspects.

\subsection{Complex eigenvalues}\label{sectCE}

From now on we will consider what happens when we abandon the assumption that $E_k$ is real for all $k$. This situation is quite interesting for concrete physical applications, both for considering some {\em effective} Hamiltonians used in different, usually non conservative, contexts, see for instance \cite{benaryeh,tripf} for an application to quantum optics, and for the discussion of broken regions in PT quantum mechanics. Once again, we assume that $H$ has $N+1$ distinct eigenvalues $E_k$, not necessarily real, corresponding to $N+1$ eigenvectors $\varphi_k$, $k=0,1,2,\ldots,N$:
\be
H\varphi_k=E_k\varphi_k.
\label{26}\en
Then  $\F_\varphi=\{\varphi_k,\,k=0,1,2,\ldots,N\}$ is automatically a basis for ${\Bbb C}^{N+1}$, admitting an unique biorthogonal basis of $\Hil$, $\F_\Psi=\{\Psi_k,\,k=0,1,2,\ldots,N\}$: $\left<\varphi_k,\Psi_l\right>=\delta_{k,l}$, for all $k, l$, and $\sum_{k=0}^N|\varphi_k\left>\right<\Psi_k|=\sum_{k=0}^N|\Psi_k\left>\right<\varphi_k|=\1$. Moreover, $H^\dagger\Psi_k=\overline{E_k}\Psi_k$ for all $k$. Of course, for all those eigenvalues $E_n$ which are real, we recover the eigenvalue equation $H^\dagger\Psi_n=E_n\Psi_n$. However, since for some $n$ $\Im(E_n)\neq0$, $H$ and $H^\dagger$ are not isospectral and we do not expect any intertwining relation as those in (\ref{23}) can be established.

As before, the operators $S_\varphi=\sum_{k=0}^N|\varphi_k\left>\right<\varphi_k|$ and $S_\Psi=\sum_{k=0}^N|\Psi_k\left>\right<\Psi_k|$ are bounded positive, self-adjoint, invertible operators, one the inverse of the other: $S_\Psi=S_\varphi^{-1}$. We can again introduce the positive square roots of $S_\Psi$ and $S_\varphi$, and we have $S_\Psi^{1/2}=S_\varphi^{-1/2}$. Also, because of the biorthogonality of $\F_\varphi$ and $\F_\psi$, we find
\be
S_\varphi\Psi_k=\varphi_k,\quad S_\Psi\varphi_k=\Psi_k,
\label{27}\en
for all $k$. Hence, so far, there are essentially no differences with respect to what we have seen in Section \ref{sectRE}, except for the expression of eigenvalue equation for $H^\dagger$, which now involves $\overline{E_k}$ rather than $E_k$. However, from now on, several differences appear, and the role of antilinear operators, see Appendix, will be essential.

To begin with, we recall that if $H\neq H^\dagger$ has only real eigenvalues, it is possible to find a bounded operator $X$, with bounded inverse, such that $H_X:=XHX^{-1}$ is self-adjoint. Equivalently, it is possible to define a new scalar product $\left<,.\right>_X$ on $\Hil$ in terms of which $H$ turns out to be self-adjoint: $\left< Hf,g\right>_X=\left< f, Hg\right>_X$. In fact, under this reality assumption, $X$ can be explicitly constructed: $X=S_\Psi^{1/2}$. Already in \cite{bagaop1} we have shown that, if at least one eigenvalue of $H$ is complex, no such scalar product exists: $H$ is {\em truly} non self-adjoint\footnote{It might be relevant to stress that here we do not mean that $H$ could be, for instance, symmetric. In this case, in fact, under suitable conditions, we could extend $H$ to a self-adjoint operator. This is not what we have in mind, of course.}! In other words, there exists no other scalar product which makes of $H$ a self-adjoint operator, with respect to its own adjoint. This is reflected also by the fact that no $X$ exists, admitting inverse, such that $H_X:=XHX^{-1}$ is self-adjoint. In fact, suppose this is not so. Then we assume for a moment that $H_X=H_X^\dagger$, but still a complex eigenvalue $E_{n_0}$ exists for $H$. Then $H\varphi_{n_0}=E_{n_0}\varphi_{n_0}$. Now, $\hat\varphi_{n_0}:=X\varphi_{n_0}$ is a non zero eigenvector of $H_X$, with $E_{n_0}$ as its eigenvalue. But this is impossible, since $H_X$ is self-adjoint.

To deal with the present situation, let us now introduce the following operators:

\be
V_\varphi f=\sum_{k=0}^N\left<f,\Psi_k\right>\varphi_k,\qquad V_\Psi f=\sum_{k=0}^N\left<f,\varphi_k\right>\Psi_k,
\label{28}\en
 for all $f\in\Hil$. It is clear that they are both antilinear: $V_\varphi (\alpha_1f_1+\alpha_2f_2)=\overline{\alpha_1}\,V_\varphi (f_1)+\overline{\alpha_2}\,V_\varphi (f_2)$ and $V_\Psi (\alpha_1f_1+\alpha_2f_2)=\overline{\alpha_1}\,V_\Psi (f_1)+\overline{\alpha_2}\,V_\Psi (f_2)$, $\forall\,\alpha_1, \alpha_2\in \Bbb C$, $\forall\,f_1, f_2\in \Hil$. Moreover
 \be
 V_\varphi^2=V_\Psi^2=\1.
 \label{29}\en
 In fact, for instance, taken $f\in\Hil$ we have
$$
V_\varphi^2 f=V_\varphi (V_\varphi f)=\sum_{k=0}^N\left<V_\varphi f,\Psi_k\right>\varphi_k=\sum_{k=0}^N\left<\left(\sum_{l=0}^N\left<f,\Psi_l\right>\varphi_l\right),\Psi_k\right>\varphi_k=
$$
$$
=\sum_{k,l=0}^N \left<\Psi_l,f\right> \left<\varphi_l,\Psi_k\right>\varphi_k=\sum_{k=0}^N \left<\Psi_k,f\right>\varphi_k=f.
$$
Here we have used the fact that $\left<\varphi_l,\Psi_k\right>=\delta_{k,l}$ and the equality $\sum_{k=0}^N|\varphi_k\left>\right<\Psi_k|=\1$.

Another useful result is that
\be
V_\varphi^\dagger=V_\Psi,
\label{210}\en
and consequently, see Appendix, $V_\Psi^\dagger=V_\varphi$. We refer to the Appendix also for the definition of the adjoint of an antilinear operator, which is slightly different from the analogous definition for linear operators. Equation (\ref{210}) follows from the following computation:
$$
\left<V_\varphi^\dagger f,g\right>=\left<V_\varphi g,f\right>=\left<\left(\sum_{k=0}^N\left<g,\Psi_k\right>\varphi_k\right),f\right>=\sum_{k=0}^N\left<\Psi_k,g\right>\left<\varphi_k,f\right>=$$ $$=\left<\left(\sum_{k=0}^N\left<f,\varphi_k\right>\Psi_k\right),g\right>=\left<V_\Psi f,g\right>,
$$
for all $f,g\in\Hil$.

Moreover, the definitions in (\ref{28}) imply that $V_\varphi\varphi_k=\varphi_k$ and $V_\Psi\Psi_k=\Psi_k$, for all $k$. If $V_\varphi$ and $V_\Psi$ were linear operators, recalling that $\dim(\Hil)<\infty$, these equalities would imply that $V_\varphi=V_\Psi=\1$, which, incidentally, would be in agreement with (\ref{29}). However, see Appendix, since  $V_\varphi$ and $V_\Psi$ are antilinear operators, this is not true. Hence, $V_\varphi$ and $V_\Psi$ are not equal to the identity operator. This is not surprising: in fact, this would imply the equality between operators of opposite nature (linear and antilinear), which is clearly never possible. This is also what happens in connection with some intertwining equations, of the kind introduced in Section \ref{sectRE}. For instance, let us consider the equality $S_\Psi H=H^\dagger S_\Psi$ in (\ref{23}). As already stated, in the present settings it is clear that this cannot be true, since it is not true already on vectors of $\F_\varphi$. In fact, while $S_\Psi H\varphi_k=E_k\Psi_k$, we have $H^\dagger S_\Psi\varphi_k=H^\dagger \Psi_k=\overline{E_k}\Psi_k$. Hence, at least for those $k$ labeling complex eigenvalues, $S_\Psi H\varphi_k\neq H^\dagger S_\Psi\varphi_k$. This fact is unpleasant, since we know that intertwining relations as those in (\ref{23}) can be extremely useful in concrete applications, as for instance in the construction of new exactly solvable models, starting from a given Hamiltonian with known eigenvectors and eigenvalues, \cite{intop,bagintop}.

A possible, apparent, way out from this enpasse can be constructed, by using the operator $V_\varphi$ introduced above. In fact, calling $H_\varphi:=V_\varphi H$, it is clear that $S_\Psi H_\varphi\varphi_k = H^\dagger S_\Psi\varphi_k$, for all $k$. However, since $H_\varphi$, and $S_\Psi H_\varphi$ as a consequence, are antilinear operators, while $H^\dagger S_\Psi$ is linear, it follows that this equality, valid for each vector of $\F_\varphi$, does not extend to all of $\Hil$. Then, $S_\Psi H_\varphi = H^\dagger S_\Psi$ is again false. However, it is not hard to show how, in fact, intertwining equations holding true in all of $\Hil$ can be deduced. The trick is simple and consists in replacing $S_\Psi H_\varphi\varphi_k = H^\dagger S_\Psi\varphi_k$ with $S_\Psi H_\varphi\varphi_k = H^\dagger V_\Psi S_\Psi\varphi_k$. This is true for all $k$, since both sides are equal to $\overline{E_k}\Psi_k$, and can be extended to all of $\Hil$, since both $S_\Psi H_\varphi$ and $H^\dagger V_\Psi S_\Psi$ are antilinear operators, now. Moreover, since $H_\varphi^\dagger = \left(V_\varphi H\right)^\dagger = H^\dagger V_\varphi^\dagger=H^\dagger V_\Psi$, our new intertwining relation becomes simply
\be
S_\Psi H_\varphi= H_\varphi^\dagger S_\Psi.
\label{211}\en
Acting of the left and from the right of this equation with $S_\varphi$ we also get another, equivalent, intertwining relation:  $H_\varphi S_\varphi= S_\varphi H_\varphi^\dagger$. The fact that $H_\varphi$ and $H_\varphi^\dagger$ are related by two (equivalent) intertwining operators suggests that they are isospectral, \cite{intop,bagintop}. In fact, this is so: $H_\varphi\varphi_n=\overline{E_n}\varphi_n$, while $H_\varphi^\dagger\Psi_n=\overline{E_n}\Psi_n$.

Due to the fact that $H$ and $V_\varphi$ do not commute in general, we can also consider a second antilinear operator $\tilde H_\varphi=HV_\varphi$, whose adjoint is $\tilde H_\varphi^\dagger=V_\Psi H^\dagger$. Of course, in general, $\tilde H_\varphi\neq \tilde H_\varphi^\dagger$ and $\tilde H_\varphi\neq H_\varphi$. Similar considerations as those discussed above for $H_\varphi$ produce the following intertwining equations for $\tilde H_\varphi$:
\be
\tilde H_\varphi S_\varphi =S_\varphi \tilde H_\varphi^\dagger,\qquad \tilde H_\varphi^\dagger S_\Psi =S_\Psi \tilde H_\varphi,
\label{212}\en
and it turns out that $\tilde H_\varphi$ and its adjoint are indeed isospectrals: $\tilde H_\varphi\varphi_n=E_n\varphi_n$, and $H_\varphi^\dagger\Psi_n=E_n\Psi_n$.

Following now what we have done in Section \ref{sectRE}, we can introduce the operators  $H_0:=S_\Psi^{1/2}H_\varphi S_\varphi^{1/2}$ and $\tilde H_0:=S_\Psi^{1/2}\tilde H_\varphi S_\varphi^{1/2}$. This can be done since both $S_\Psi$ and $S_\varphi$ admit unique positive square root. Moreover, no problem with domains arises, in the present context, since $\Hil$ is finite dimensional. Of course, both these operators are antilinear. Moreover, they are both self-adjoint: $H_0=H_0^\dagger$ and $\tilde H_0=\tilde H_0^\dagger$.

Let now put $e_k=S_\Psi^{1/2}\varphi_k= S_\varphi^{1/2}\Psi_k$, $k=0,1,2,\ldots,N$,  and  $\E=\{e_k,\,k=0,1,2,\ldots,N\}$. It is easy to check that these vectors are eigenstates of both $H_0$ and $H_0^\dagger$: $H_0e_k=\overline{E_k}\, e_k$ and $\tilde H_0e_k=E_k\, e_k$, for all $k$. As discussed in the Appendix, the fact that eigenvalues of a self-adjoint, antilinear, operators can be complex should not be a surprise. On the other hand, and this was not granted, see Appendix, the different $e_k$'s here are still indeed orthogonal: $$\left<e_k,e_n\right>=\left<S_\Psi^{1/2}\varphi_k,S_\Psi^{1/2}\varphi_n\right>=\left<S_\Psi\varphi_k,\varphi_n\right>=\left<\Psi_k,\varphi_n\right>=\delta_{k,n}.$$
Hence $\E$ is an o.n. basis for $\Hil$.

\vspace{2mm}

The intertwining relations we have deduced so far, in presence of complex eigenvalues of $H$, involve antilinear operators. In fact, this seems the only way to deduce relations between some operator related to the original Hamiltonian of the system and its adjoint. However, it is also possible to deduce intertwining relations between linear operators. In fact, simple computations allow us to deduce that
\be
S_\Psi H_{\varphi,\varphi}=H^\dagger S_\Psi,
\label{213}\en as well as the equivalent equalities $H_{\varphi,\varphi} S_\varphi=S_\varphi H^\dagger$, $S_\varphi H_{\varphi,\varphi}^\dagger=HS_\varphi$ and $S_\Psi H= H_{\varphi,\varphi}^\dagger S_\Psi$. Here $H_{\varphi,\varphi}=V_\varphi H V_\varphi$, which is linear. Hence, these equalities all involve just linear operators. For this reason, from one side, they might appear  more interesting than those deduced before. However, as stated, what makes (\ref{213}) possibly less interesting for us, is that it involves not really $H$ and its adjoint, or $H_{\varphi,\varphi}$ and its adjoint, but both $H_{\varphi,\varphi}$ and $H^\dagger$. On the other hand, equation (\ref{211}) just involves $H_\varphi$ and its adjoint, which is closer to what usually happens in PT-quantum mechanics and in the theory of intertwining operators.

\vspace{2mm}

As in Section \ref{sectRE}, using $S_\varphi$ and $S_\Psi$ we can introduce different scalar products as in (\ref{24}), $\left<.,.\right>_\varphi$ and $\left<.,.\right>_\Psi$, and the adjoints associated to these, $\flat$ and $\sharp$, and this can be done, following the recipe given in the Appendix, both for linear and for antilinear operators. The same relations as those deduced in (\ref{2add1}) can be defined for both these kind of operators. In particular, we deduce that
\be
H_\varphi=H_\varphi^\sharp,\quad H_\varphi^\dagger=\left(H_\varphi^\dagger\right)^\flat,\quad \tilde H_\varphi=\tilde H_\varphi^\sharp,\quad \tilde H_\varphi^\dagger=\left(\tilde H_\varphi^\dagger\right)^\flat.
\label{214}\en
Hence, even if $H_\varphi$ and $\tilde H_\varphi$ are antilinear, they still obey the same self-adjoint properties as their linear counterparts. We will not investigate this aspect of these operators here, since this is not relevant for what we are doing in this paper and because this does not appear to have useful consequences, in concrete applications. Rather than this, we would like to point out that, as it will be clear also in the next section, going from the unbroken region (UR) to the broken region (BR) appears as a sort of phase transition: in the UR, characterized by $H$ having only real eigenvalues, the time evolution, for models living in finite dimensional Hilbert spaces, is necessarily periodic or quasi-periodic, depending on the mutual ratios of the various real eigenvalues of the Hamiltonian $H$. No damping is possible, and the transition probabilities appear to be also periodic or quasi-periodic in time, \cite{bagaop1}. In the BR, in which some eigenvalues are surely complex, the situation is completely different, and non trivial asymptotic behaviors can be deduced, while periodicity (or quasi-periodicity) is lost. This will be made clear in Section \ref{sectTPandC}.

\vspace{2mm}

{\bf Remark:--} We should mention that the connection of pseudo-hermitian operators and antilinear operators, or more exactly with antilinear symmetries, has already been considered in the context of non self-adjoint Hamiltonians by several authors, see for instance A. Mostafazadeh in \cite{am} and L. Solombrino in \cite{sol}, but with a different perspective with respect to ours.

\section{An example}\label{sectAnex}

As an explicit application of our general settings we consider the following two-by-two matrix:
\be
H=\frac{1}{2-\alpha\beta}\left(
                           \begin{array}{cc}
                             2E_1-\alpha\beta E_2 & \alpha(E_2-E_1) \\
                             2\beta(E_1-E_2) & 2E_2-\alpha\beta E_1 \\
                           \end{array}
                         \right),
                         \label{31}\en
where $\alpha$ and $\beta$ are real parameters such that $\alpha\beta\neq2$, while $E_1$ and $E_2$ are, in general, complex quantities. Of course, without further assumptions on these quantities, $H\neq H^\dagger$. This is the situation we will consider here. It is easy to find the eigenstates and the eigenvalues of $H$ and $H^\dagger$:
\be
\varphi_1=\left(
            \begin{array}{c}
              1 \\
              \beta \\
            \end{array}
          \right), \quad \varphi_2=\left(
            \begin{array}{c}
              \alpha \\
              2 \\
            \end{array}
          \right),\quad
\Psi_1=\frac{1}{2-\alpha\beta}\left(
            \begin{array}{c}
              2 \\
              -\alpha \\
            \end{array}
          \right),\quad
          \Psi_2=\frac{1}{2-\alpha\beta}\left(
            \begin{array}{c}
              -\beta \\
              1 \\
            \end{array}
          \right)
          \label{32}\en
With this normalization we find that
\be
\left<\varphi_k,\Psi_n\right>=\delta_{k,n},
\label{33}\en
while both $\left<\varphi_1,\varphi_2\right>$ and $\left<\Psi_1,\Psi_2\right>$ are different from zero. Hence the two sets $\F_\varphi=\{\varphi_1,\varphi_2\}$ and $\F_\Psi=\{\Psi_1,\Psi_2\}$ are biorthogonal bases for $\Hil=\Bbb C^2$: each $f\in\Hil$ can be written as $f=\sum_{j=1}^2\left<\Psi_j,f\right>\varphi_j = \sum_{j=1}^2\left<\varphi_j,f\right>\Psi_j$. In bra-ket language, $\F_\varphi$ and $\F_\Psi$ produce the following resolutions of the identity: $\1=\sum_{j=1}^2|\Psi_j\left>\right<\varphi_j|=\sum_{j=1}^2|\varphi_j\left>\right<\Psi_j|$.

Moreover:
\be
H\varphi_j=E_j\varphi_j, \qquad H^\dagger\Psi_j=\overline{E_j}\Psi_j,
\label{34}\en
$j=1,2$. Hence, the parameters $E_1$ and $E_2$ in $H$, see (\ref{31}), are exactly its eigenvalues, while their complex conjugates are the eigenvalues of $H^\dagger$.

It is now easy to check that $V_\varphi=V_\Psi=V$, the conjugation antilinear operator acting as follows: $V\left(
                                                                                                               \begin{array}{c}
                                                                                                                 f_1 \\
                                                                                                                 f_2 \\
                                                                                                               \end{array}
                                                                                                             \right)=\left(
                                                                                                               \begin{array}{c}
                                                                                                                 \overline{f_1} \\
                                                                                                                 \overline{f_2} \\
                                                                                                               \end{array}
                                                                                                             \right)
$. $V$ is self-adjoint: $V=V^\dagger$, and satisfies, clearly, the equality $V^2=\1$, as expected.

In order to check the intertwining relations deduced in Section \ref{sectCE}, we need first to compute $S_\varphi$ and $S_\Psi$, which turn out to be
$$
S_\varphi=\left(
            \begin{array}{cc}
              1+\alpha^2 & \beta+2\alpha \\
              \beta+2\alpha & 4+\beta^2 \\
            \end{array}
          \right), \quad
          S_\Psi=\frac{1}{(2-\alpha\beta)^2}\left(
            \begin{array}{cc}
              4+\beta^2 & -(\beta+2\alpha) \\
              -(\beta+2\alpha) & 1+\alpha^2 \\
            \end{array}
          \right).
$$
These matrices are both manifestly self-adjoint. Moreover, they are also positive and one is the inverse of the other: $S_\varphi=S_\Psi^{-1}$. Now, equalities (\ref{211}) and (\ref{212}) can be explicitly deduced. The computations are not difficult, and will not be given here. More interesting is to see what is the expression of the linear operator $H_{\varphi,\varphi}=VHV$ in (\ref{213}), and how this is related to the original Hamiltonian $H$. In fact, it turns out that
$$
H_{\varphi,\varphi}=\frac{1}{2-\alpha\beta}\left(
                           \begin{array}{cc}
                             2\overline{E_1}-\alpha\beta \overline{E_2} & \alpha(\overline{E_2}-\overline{E_1}) \\
                             2\beta(\overline{E_1}-\overline{E_2}) & 2\overline{E_2}-\alpha\beta \overline{E_1} \\
                           \end{array}
                         \right),$$
which shows that $H_{\varphi,\varphi}$ differs from $H$ only because each $E_j$ is replaced by its complex conjugate. Of course, this means, in particular, that $H_{\varphi,\varphi}=H$ if $E_1$ and $E_2$ are real. It is now straightforward to check all the other results and equalities discussed in Section \ref{sectCE}.

\vspace{2mm}

It is interesting to notice that the Hamiltonian $H$ in (\ref{31}) is  strongly related to the Hamiltonian
$$
h=\left(
    \begin{array}{cc}
      r e^{i\theta} & s e^{i\Phi} \\
      t e^{-i\Phi} & r e^{-i\theta} \\
    \end{array}
  \right),
$$
originally introduced in \cite{das} in connection with $PT$-quantum mechanics. Moreover, it is known that $h$ extends a similar one, with $t=s$, already considered, for instance, in \cite{ben1}. Notice that in $h$ all the parameters, $r,s,t,\theta$ and $\Phi$, are real. Since we are interested in the so called BR, in which the eigenvalues of $h$ turn out to be complex conjugate, we need to require that $r^2\sin^2(\theta)-ts>0$, which is the case we  consider here. In fact, in the UR, the eigenvalues are real, and the framework discussed in this paper is not particularly relevant.

If $\Phi=-\frac{\pi}{2}$, $H$ and $h$ coincide if we take $\alpha$, $\beta$, $E_1$ and $E_2$ as follows:
$$
\beta=\frac{1}{s}\left(r\sin(\theta)\pm\sqrt{r^2\sin^2(\theta)-ts}\right), \quad \alpha=2\beta\,\frac{s}{t},
$$
while, calling $R=\Re(E_1)=\Re(E_2)$ and $I=\Im(E_1)=-\Im(E_2)$, we put
$$
R=r\cos(\theta), \quad I=\frac{2-\alpha\beta}{2\alpha}\,s.
$$
In particular we see that these formulas confirm that $\alpha$ and $\beta$ are both real. Moreover, $2-\alpha\beta$ is always non zero if $s$ and $t$ are non zero.

This simple example already shows that our general settings can be applied to models already considered in the literature, and can be useful to study exceptional points and possible transitions from unbroken to broken regions.

\subsection{Transition probabilities}\label{sectTPandC}

In two recent papers, \cite{bagaop1,bagaop2}, different definitions of the dynamics for quantum systems driven by non-self-adjoint Hamiltonians, and consequently for transition probabilities, have been considered and compared. The possibility of considering different definitions of transition probabilities is of course related to the existence of various scalar products in the same Hilbert space. However, in \cite{bagaop2} we have seen that, if the system lives in an infinite-dimensional Hilbert space $\Lc^2(\Bbb R)$, the appropriate choice of scalar product seems to be the  {\em standard} one, $\left<f,g\right>=\int_{\Bbb R}\overline{f(x)}\,g(x)\,dx$. Otherwise, as discussed in \cite{bagaop2} in connection with the Swanson model, the range of the parameters defining the original model should be restricted to keep the model well-defined during its time evolution. Driven by this idea we restrict here, for the model described by the Hamiltonian in (\ref{31}), to
the ordinary scalar product in $\Bbb C^2$ and to the following definition of the transition probability of going from a state $\Phi_0$ to a state $\Phi_f$ at time $t$:
\be
P_{\Phi_0\rightarrow\Phi_f}(t)=\frac{\left|\left<\Phi_f, \Phi(t)\right>\right|^2}{\|\Phi_f\|^2\|\Phi(t)\|^2},
\label{35}\en
where $\Phi(t)=e^{-iHt}\Phi_0$. We will now briefly discuss what happens, with this definition, when going from the unbroken to the broken region, for a initial vector $\Phi_0=c_1\varphi_1+c_2\varphi_2$ and a final vector $\Phi_f=d_1\Psi_1+d_2\Psi_2$.  Using both $\F_\varphi$ and $\F_\Psi$ to expand the initial and the final states is mathematically correct (since they are both bases) and technically convenient (since they are biorthogonal), even if it may appear not entirely natural.

Simple computations produce, first of all,
$$
\|\Phi_f\|^2=\frac{1}{(2-\alpha\beta)^2}\left(|d_1|^2(4+\alpha^2)+|d_2|^2(1+\beta^2)-(\alpha+2\beta)(\overline{d_1}d_2+\overline{d_2}d_1)\right),
$$
which does not depend on $E_1$ and $E_2$. Then, $\|\Phi_f\|^2$ does not change changing region, i.e. going from the UR to the BR. On the other hand, in the UR ($E_1$ and $E_2$ real) we find
$$
\|\Phi(t)\|^2=|c_1|^2(1+\beta^2)+|c_2|^2(4+\alpha^2)+(\alpha+2\beta)\left(\overline{c_1}c_2e^{i(E_1-E_2)t}+c_1\overline{c_2}e^{-i(E_1-E_2)t}\right),
$$
and
$$
\left|\left<\Phi_f, \Phi(t)\right>\right|^2=|c_1|^2|d_1|^2+|c_2|^2|d_2|^2+
\left(\overline{c_1}c_2d_1\overline{d_2}e^{i(E_1-E_2)t}+c_1\overline{c_2}\overline{d_1}d_2e^{-i(E_1-E_2)t}\right).
$$
In the BR  ($E_1=\overline{E_2}=R+iI$) these results must be replaced by the following formulas
$$
\|\Phi(t)\|^2=|c_1|^2(1+\beta^2)e^{2It}+|c_2|^2(4+\alpha^2)e^{-2It}+(\alpha+2\beta)\left(\overline{c_1}c_2+c_1\overline{c_2}\right),
$$
and
$$
\left|\left<\Phi_f, \Phi(t)\right>\right|^2=|c_1|^2|d_1|^2e^{2It}+|c_2|^2|d_2|^2e^{-2It}+
\left(\overline{c_1}c_2d_1\overline{d_2}+c_1\overline{c_2}\overline{d_1}d_2\right).
$$
It is clear that in the case of UR, only oscillations are possible, with a period $T=\frac{2\pi}{E_1-E_2}$. On the other hand, in case of BR, no non-trivial oscillation is possible. For example, if $\Phi_0=\varphi_1+\varphi_2$ and $\Phi_f=\Psi_1$, in the BR we find
$$
P_{\Phi_0\rightarrow\Phi_f}(t)=\frac{(2-\alpha\beta)^2\,e^{2It}}{(4+\alpha^2)\left((1+\beta^2)e^{2It}+(4+\alpha^2)e^{-2It}+2(\alpha+2\beta)\right)},
$$
which tends to $\frac{(2-\alpha\beta)^2}{(4+\alpha^2)(1+\beta^2)}$ when $t\rightarrow\infty$ if $I>0$, or to zero if $I<0$. So, as already stated, we get two completely different behaviors depending on the region in which our system is, broken or unbroken\footnote{We do not expect significant differences, if not for some minor aspects, if we consider, rather than $P_{\Phi_0\rightarrow\Phi_f}(t)$, the other transition functions $P_{\Phi_0\rightarrow\Phi_f}^\varphi(t)$ and $P_{\Phi_0\rightarrow\Phi_f}^\Psi(t)$ introduced in \cite{bagaop1}, using the scalar products introduced in (\ref{24}).}. This could be useful in some concrete application, for instance in quantum optics: depending on the region considered we can use the same Hamiltonian to describe oscillations or damping, which can be relevant to describe decay phenomena occurring for particular range of values of the parameters of the model.

\section{Conclusions}

We have seen how the presence of complex eigenvalues in the spectrum of a given Hamiltonian changes drastically both the mathematical settings, causing the appearance of antilinear operators, and the physical conclusions, leading to non periodic transition probabilities. We have deduced intertwining relations similar to those already found in \cite{bagaop1}, and intertwining relations of a different kind, involving two different, although related, operators.

We have also discussed the possibility of using a single underlying framework to describe real or complex eigenvalues, and their relevance in connection with systems behaving in a different way depending on the values of the parameters.

\section*{Acknowledgements}
The author would like to acknowledge  support from the
   Universit\`a di Palermo and from Gnfm.

 \renewcommand{\theequation}{A.\arabic{equation}}

 \section*{Appendix:  few useful facts on antilinear operators}

Most of the results on operators which can be found in the literature refers to linear ones. In this paper we have seen, however, that the set of the linear operators is not sufficient to deal with Hamiltonians as the ones considered here. For this reason, also in view of the fact that the literature on antilinear operator is not particularly reach, we have decided to list in this appendix a set of properties of these operators which are used in the main part of this paper. We also refer to \cite{uhl} for a very recent review on this subject, and to \cite{herbut} for an older, very useful, paper.

Let $\Hil={\Bbb C}^{N+1}$ be our finite-dimensional Hilbert space, with scalar product $\left<.,.\right>$. A map $V:\Hil\rightarrow\Hil$ is said to be {\em antilinear} if $V(a_1\varphi_1+a_2\varphi_2)=\overline{a_1 }\varphi_1+\overline{a_2 }\varphi_2$, for all $a_1, a_2\in \Bbb C$ and for all $\varphi_1, \varphi_2\in\Hil$.   We call $\Lc(\Hil)$ and $\Lc_{al}(\Hil)$ respectively the  set of all the linear and of all the antilinear operators on $\Hil$.

Some facts can be easily proved:

\begin{enumerate}

\item If $A\in\Lc(\Hil)$ and $V\in\Lc_{al}(\Hil)$, then $AV$ and $VA$ are both antilinear. They act on $\varphi\in\Hil$ as follows: $(AV)\varphi=A(V\varphi)$, and $(VA)\varphi=V(A\varphi)$. $A+V$ is neither linear nor antilinear.

\item If $V_1, V_2\in \Lc_{al}(\Hil)$ then $V_1V_2\in \Lc(\Hil)$.

\end{enumerate}

An interesting difference between linear and antilinear operators is the following: let $\F_\varphi=\{\varphi_0,\ldots,\varphi_{N}\}$ be a (not necessarily o.n.) basis for $\Hil$. Let $A\in\Lc(\Hil)$ be such that $A\varphi_j=\varphi_j$, for all $j$. Hence, it is well known that $A=\1$, $\1$ being the identity operator on $\Hil$. On the other hand, let $V\in\Lc_{al}(\Hil)$ be such that  $V\varphi_j=\varphi_j$, for all $j$. Then, it is easy to see that $V\neq\1$. This is, in fact, not surprising, since otherwise we would have a linear operator ($\1$) equal to an antilinear operator ($V$). This is clearly impossible. In fact, using the definition of antilinearity, we have $V(i\varphi_j)=-iV(\varphi_j)=-i\varphi_j\neq i\varphi_j=\1(i\varphi_j)$. Hence, as stated, $V$ must be different from the identity operator and, in fact, we have already introduced in Section \ref{sectAnex} an antilinear operator $V$ satisfying $V\varphi_j=\varphi_j$, $j=1,2,3$, but clearly different from the identity operator. This result is quite general: it might happen, as it happens in Section \ref{sectCE}, that a linear operator $C$ and an antilinear operator $W$ act in the same way on a basis $\F_\varphi$ of $\Hil$: $C\varphi_j=W\varphi_j$, for all $j$. However, for the same reason which allows us to conclude that $V\neq\1$, this does not imply that $C=W$. On the other hand, if $A, C\in\Lc(\Hil)$ and $V, W\in\Lc_{al}(\Hil)$ are such that $A\varphi_j=C\varphi_j$ and $V\varphi_j=W\varphi_j$, for all $j$, then $A=C$ and $V=W$.

Another difference between linear and antilinear operators concerns the definition of the adjoint. If $A\in \Lc(\Hil)$, it is well known that $A^\dagger$ is defined by $\left<A^\dagger \varphi,\phi\right>=\left<\varphi,A\phi\right>$, for all $\varphi,\phi\in\Hil$. On the other hand, if we try to define the adjoint of $V\in\Lc_{al}(\Hil)$ in the same way, we loose the antilinearity for $V^\dagger$. This is preserved adopting the following definition:
\be\label{a1}
\left<V^\dagger \varphi,\phi\right>=\overline{\left<\varphi,V\phi\right>}=\left<V\phi,\varphi\right>,
\en
for all $\varphi,\phi\in\Hil$. In fact, with this definition, $\Lc_{al}(\Hil)$ turns out to be closed under adjoint: if $V\in\Lc_{al}(\Hil)$, then $V^\dagger\in\Lc_{al}(\Hil)$ as well.
Notice also that $(V^\dagger)^\dagger=V$, and that $(V_1V_2)^\dagger=V_2^\dagger V_1^\dagger$, for all $V, V_1, V_2\in\Lc_{al}(\Hil)$, so that the usual properties of the adjoint map are recovered also in this case.

We say that the antilinear operator $V$ is self-adjoint if $V=V^\dagger$. For linear operators, this requirement has many important consequences: if $A=A^\dagger\in\Lc(\Hil)$ admits  eigenvalues $\lambda_1, \lambda_2, \ldots$, these are all real. Moreover, if $\psi_1$ and $\psi_2$ are the eigenvectors of $A$ corresponding to $\lambda_1$ and $\lambda_2$, $A\psi_j=\lambda_j\psi_j$, $j=1,2$, and if $\lambda_1\neq\lambda_2$, then $\left<\psi_1,\psi_2\right>=0$. Both these properties are false for antilinear operators. For instance, let $\psi$ be a normalized vector in $\Hil$, $\alpha$ a fixed complex number, with $\Im(\alpha)\neq0$, and $V$ be the antilinear operator defined as $Vf=\alpha\left<f,\psi\right>\psi$. Using (\ref{a1}) one can check that $V$ is self-adjoint: $V=V^\dagger$. Moreover, $\psi$ is an eigenstate of $V$ with complex eigenvalue $\alpha$: $V\psi=\alpha\psi$.

Let us now introduce $z_1=e^{i\frac{\pi}{4}}$ and $z_2=e^{i\frac{\pi}{3}}$, and two proportional vectors $\psi_{z_j}=z_j\psi$, $j=1,2$. It is easy to check that these two vectors are eigenstates of $V$ corresponding to different eigenvalues: $V\psi_{z_j}=\alpha_{j}\psi_{z_j}$, $j=1,2$, where $\alpha_{1}=-i\alpha$ and $\alpha_{2}=-\frac{\alpha}{2}(1+i\sqrt{3})$. However, as it is clear, $\left<\psi_{z_1},\psi_{z_2}\right>=e^{i\frac{\pi}{12}}\neq0$. 

The same conclusion can be deduced as follows\footnote{I wish to thank the Referee for suggesting me this example.}: let $\psi$ be an eigenstate of $V$ with complex eigenvalue $\alpha$: $V\psi=\alpha\psi$, let $\lambda$ be a complex number with $\Im(\lambda)\neq0$ and let also $\Phi=\lambda\psi$. Then
$$
V\Phi=V\left(\lambda\psi\right)=\overline{\lambda}(V\psi)=\overline{\lambda}(\alpha\psi)=\left(\frac{\overline{\lambda}}{\lambda}\,\alpha\right)\Phi.
$$ 
Of course, since $\overline{\lambda}\neq\lambda$, $\Phi$ and $\psi$ are both eigenstates of $V$ corresponding to different eigenvalues. However $\left<\psi,\Phi\right>=\lambda\|\psi\|^2\neq0$.

It is important to notice that, in both cases above example, the eigenvalues are different but have the same modulus. This is important. In fact, let us consider the eigenvalue equation $V\varphi_j=E_j\varphi_j$, $j=1,2$, for some $V\in\Lc_{al}(\Hil)$, non zero $\varphi_j\in\Hil$ and complex quantities $E_j$ with $|E_1|\neq|E_2|$. Then, it is easy to check that, in this case, orthogonality is still satisfied: $\left<\varphi_1,\varphi_2\right>=0$. This follows from the fact that $\varphi_j$ is also eigenstate of the linear, self-adjoint operator $V^2$ with eigenvalue $|E_j|^2$: $V^2\varphi_j=V(E_j\varphi_j)=\overline{E_j}V\varphi_j=|E_j|^2\varphi_j$, and from the assumption that $|E_1|^2\neq|E_2|^2$. Hence, $\varphi_1$ and $\varphi_2$ are also eigenstates of the self-adjoint operator $V^2$ corresponding to different (real) eigenvalues; then they are orthogonal.


\begin{thebibliography}{99}


\bibitem{baginbagbook} F. Bagarello, {\em Deformed canonical (anti-)commutation relations and non hermitian hamiltonians}, in {Non-selfadjoint operators in quantum physics: Mathematical aspects}, F. Bagarello, J. P. Gazeau, F. H. Szafraniek and M. Znojil Eds., John Wiley and Sons Eds., to appear in April 2015.

\bibitem{ben}C. Bender, {\em Making Sense of Non-Hermitian Hamiltonians}, Rep. Progr.  Phys., {\bf 70},  947-1018 (2007)


\bibitem{ali} A. Mostafazadeh, {\em Pseudo-hermitian representation of quantum mechanics},  Int. J. Geom. Methods Mod. Phys., {\bf 7}, 1191-1306 (2010)

\bibitem{bagaop1} F. Bagarello, {\em Some results on the dynamics and  transition probabilities for non self-adjoint hamiltonians},  Ann. of Phys., {\bf 356}, 171-184 (2015)

\bibitem{bagaop2} F. Bagarello, {\em Transition probabilities for non self-adjoint Hamiltonians in infinite dimensional Hilbert spaces},
Ann. of Phys., in press

\bibitem{you} R. M. Young, {\em On complete biorthogonal bases}, Proceedings of the American Mathematical Society, {\bf 83}, No. 3, 537-540, (1981)

\bibitem{chri} O. Christensen, {\em An Introduction to Frames and Riesz Bases}, Birkh\"auser, Boston, (2003)

\bibitem{benaryeh} Y. Ben-Aryeh, A. Mann, I. Yaakov, {\em Rabi oscillations in a two-level atomic system with a pseudo-hermitian hamiltonian},
J. Phys. A, {\bf 37} 12059-12066, (2004)


\bibitem{tripf} O. Cherbal, M. Drir, M. Maamache , D. A. Trifonov, {\em Fermionic coherent states for pseudo-Hermitian two-level systems},
J. Phys. A, {\bf 40}, 1835-1844, (2007)




\bibitem{intop} Kuru S., Tegmen A., Vercin A., {\em Intertwined isospectral potentials in an arbitrary dimension},
J. Math. Phys, {\bf 42}, No. 8, 3344-3360, (2001); Kuru S.,
Demircioglu B., Onder M., Vercin A., {\em Two families of
superintegrable and isospectral potentials in two dimensions}, J.
Math. Phys, {\bf 43}, No. 5, 2133-2150, (2002); Samani K. A., Zarei
M., {\em Intertwined hamiltonians in two-dimensional curved spaces},
Ann. of Phys., {\bf 316}, 466-482, (2005); N. Aizawa, V. K. Dobrev, {\em Intertwining Operator Realization of Non-Relativistic Holography}, Nucl. Phys. B {\bf 828}, 581-593 (2010); B. Midya, B. Roy, R. Roychoudhury, {\em Position Dependent Mass Schroedinger Equation and Isospectral Potentials : Intertwining Operator approach}, J.  Math. Phys., {\bf 51}, 022109 (2010); A. L. Lisok, A. V. Shapovalov, A. Yu. Trifonov, {\em Symmetry and Intertwining Operators for the Nonlocal Gross-Pitaevskii Equation}, SIGMA {\bf 9}, 066, 21 pages (2013)


\bibitem{bagintop} F. Bagarello {\em Extended SUSY quantum mechanics, intertwining operators and coherent states},
  Phys. Lett. A,  {\bf 372}, 6226-6231 (2008); F. Bagarello {\em Vector coherent states and intertwining operators},
   J. Phys. A., doi:10.1088/1751-8113/42/7/075302, {\bf 42} (2009) 075302 (11pp); F. Bagarello, {\em Intertwining operators between different Hilbert spaces: connection with frames}, J. Math. Phys.,  {\bf 50}, 043509 (2009) (13pp); F. Bagarello {\em Quons, coherent states and intertwining operators}, Phys. Lett.  A, {\bf 373}, 2637-2642 (2009); F. Bagarello {\em Mathematical aspects of intertwining
operators: the role of Riesz bases},   J. Phys. A,  {\bf 43},  175203 (2010) (12pp); F. Bagarello, {\em Non isospectral hamiltonians, intertwining operators and hidden hermiticity}, Phys. Lett. A, {\bf 376}, 70-74 (2011)


\bibitem{am} A. Mostafazadeh, {\em Pseudo-Hermiticity versus PT Symmetry III: Equivalence of pseudo-Hermiticity and the presence of antilinear symmetries},  J. Math. Phys. {\bf 43}, 3944-3951 (2002)


\bibitem{sol} L. Solombrino, {\em Weak pseudo-Hermiticity and antilinear commutant},  J. Math. Phys. {\bf 43}, 5439 (2002)





\bibitem{das} A. Das, L. Greenwood {\em An alternative construction of the positive inner product for pseudo-Hermitian Hamiltonians: examples}, J. Math. Phys., {\bf 51}, Issue 4, 042103 (2010)




\bibitem{ben1} C. M. Bender, M. V. Berry, A. Mandilara, {\em Generalized PT Symmetry and Real Spectra}, J.  Phys. A, {\bf 35}, L467, (2002)


\bibitem{uhl} A. Ulhmann, {\em Anti- (conjugate) linearity},  arXiv:1507.06545


\bibitem{herbut} F. Herbut, M. Vujicic, {\em Basic algebra of antilinear operators and some applications. I}, J. Math. Phys. {\bf 8}, No. 6, 1345-1354 (1967)















\end{thebibliography}
\end{document}